\documentclass{amsart}
\usepackage{amssymb}
\usepackage{latexsym}

\newcommand{\ZZ}{{\mathbb Z}}
\newcommand{\RR}{{\mathbb R}}
\newcommand{\CC}{{\mathbb C}}
\newcommand{\NN}{{\mathbb N}}

\newtheorem{theorem}{Theorem}

\newtheorem{lemma}{Lemma}[section]
\newtheorem{prop}[lemma]{Proposition}
\newtheorem{coro}[lemma]{Corollary}

\renewcommand{\Im}{{\rm Im}}

\newcommand{\tr}{{\mathrm{tr}}}

\newcounter{smalllist}

\begin{document}
\title{Log-dimensional spectral properties of one-dimensional quasicrystals}
\author[D.~Damanik, M.~Landrigan]{David Damanik$\, ^1$ and Michael Landrigan$\, ^2$}
\thanks{D.\ D.\ was supported in part by the National Science Foundation through Grant DMS--0010101}
\maketitle
\vspace{0.3cm}
\noindent
$^1$ Department of Mathematics 253--37, California Institute of Technology, Pasadena, CA 91125, USA\\[2mm]
$^2$ Department of Mathematics, Idaho State University, Pocatello, ID 83209\\[3mm]
E-mail: \mbox{damanik@its.caltech.edu, landmich@isu.edu}\\[3mm]

\begin{abstract}
We consider discrete one-dimensional Schr\"odinger operators on the whole line and establish a criterion for continuity of spectral measures with respect to $\log$-Hausdorff measures. We apply this result to operators with Sturmian potentials and thereby prove logarithmic quantum dynamical lower bounds for all coupling constants and almost all rotation numbers, uniformly in the phase.
\end{abstract}

\section{Introduction}

We are interested in discrete one-dimensional Schr\"odinger operators in $\ell^2(\ZZ)$ given by

\begin{equation}\label{oper}
(H\phi)(n) =  \phi(n+1) + \phi(n-1) + V(n)\phi(n)
\end{equation}
with potential $V\!\!:\!\ZZ\to\RR$. To each such whole-line operator we associate two half-line operators, $H_+=P_+^* H P_+$ and $H_-=P_-^* H P_-$, where $P_\pm$ denote the inclusions
$P_+\!:\!\ell^2(\{1,2,...\})\hookrightarrow\ell^2(\ZZ)$ and
$P_-\!:\!\ell^2(\{0,-1,-2,...\})\hookrightarrow\ell^2(\ZZ)$.

For each $z\in\CC\setminus\RR$ we define $\psi^\pm(n;z)$ to be the unique solutions to the difference equation

\begin{equation}\label{eve}
\psi(n+1) + \psi(n-1) + V(n) \psi(n) = E \psi(n)
\end{equation}
with
$$
\psi^\pm(0;z)=1\quad\text{and}\quad\sum_{n=0}^\infty |\psi^\pm(\pm n;z) |^2 < \infty.
$$
With this notation we can define the Weyl functions by
\begin{align*}
m^+(z) &= \langle\delta_1 | (H_+-z)^{-1} \delta_1 \rangle = -\psi^+(1;z)/\psi^+(0;z)\\
m^-(z) &= \langle\delta_0 | (H_--z)^{-1} \delta_0 \rangle = -\psi^-(0;z)/\psi^-(1;z)
\end{align*}
for each $z\in\CC\setminus\RR$. Here and elsewhere, $\delta_n$ denotes the vector in $\ell^2$ supported at $n$ with $\delta_n(n)=1$. For the whole-line problem, the $m$-function role is played by the $2\times2$ matrix $M(z)$:
$$
\left[\begin{smallmatrix}  a \\  b \end{smallmatrix}\right]^\dagger M(z)
\left[\begin{smallmatrix}  a \\  b \end{smallmatrix}\right]
= \big\langle (a\delta_0+b\delta_1)\big|(H-z)^{-1} (a\delta_0+b\delta_1) \big\rangle.
$$
Or, more explicitly,
\begin{align*}
M &= \frac{1}{\psi^+(1)\psi^-(0)-\psi^+(0)\psi^-(1)}
\begin{bmatrix} \psi^+(0)\psi^-(0) & \psi^+(1)\psi^-(0) \\
\psi^+(1)\psi^-(0) & \psi^+(1)\psi^-(1) \end{bmatrix} \\
&= \frac{1}{1-m^+m^-}
\begin{bmatrix} m^- & -m^+m^- \\  -m^+m^- & m^+ \end{bmatrix}
\end{align*}
with $z$ dependence suppressed. We define $m(z)=\tr\big(M(z)\big)$, that is, the trace of $M$. These definitions relate the $m$-functions to resolvents and hence to spectral measures. By pursuing these relations, one finds that:

\begin{align}
\notag
m^{\pm}(z) &= \int \frac{1}{t-z} d\rho^{\pm}(t), \\
\label{4:mRep}
m(z) &= \int \frac{1}{t-z} d\Lambda(t),
\end{align}
where $\rho^{+},\rho^{-}$ are the spectral measures for the pairs $(H_+,\delta_1),(H_-,\delta_0)$, respectively, and $\Lambda$ is the sum of the spectral measures for the pairs $(H,\delta_0)$ and $(H,\delta_1)$. It is known that the pair of vectors $\{\delta_0,\delta_1\}$ is cyclic for $H$.

Our goal is to find a criterion for $\Lambda$ to be absolutely continuous with respect to logarithmic Hausdorff measures. Let us first recall the notion of Hausdorff measure, and logarithmic Hausdorff measure in particular. Given a function $h : [0, \infty) \rightarrow [0,\infty)$ which is continuous with $h(0)=0$, a so-called dimension function, define for $S$ a subset of $\RR$,
$$
\mu_h(S) = \lim_{\delta \rightarrow 0} \inf_{\delta-{\rm covers}} \sum_{i=1}^\infty h(b_i - a_i),
$$
where a $\delta$-cover is a cover of $S$ by intervals $(a_i,b_i)$, $i \in \NN$ of length at most $\delta$. When restricted to Borel sets, this gives rise to a measure $\mu_h$, called Hausdorff measure corresponding to the dimension function $h$. For example, $h(x) = x^\alpha$, $0 < \alpha < 1$ or $h(x) = \log b (x) = (\log \frac{1}{x})^{-b}$, $b > 0$. For Hausdorff measures $\mu_{x^\alpha}$, a criterion for absolute continuity of $\Lambda$ was found in \cite{dkl}. This criterion is based on power law upper and lower bounds for solutions to \eqref{eve}
and the proof uses the Jitomirskaya-Last \cite{jl1} extension of Gilbert-Pearson theory \cite{gp}. We will prove a similar criterion for absolute continuity of $\Lambda$ with respect to $\mu_{\log b}$ which is based only on power law lower bounds for solutions of \eqref{eve}. This is motivated by our application of this criterion to operators with Sturmian potentials where the lower bounds can be shown for almost all rotation numbers, whereas upper bounds are known only for a zero-measure set of rotation numbers.

Explicitly, we will show the following:

\begin{theorem}\label{crit}
Suppose $V$ is bounded and there is $\gamma > 0$ such that for $\Lambda$-almost every energy $E$, every solution of \eqref{eve} with

\begin{equation}\label{normal}
|\psi(0)|^2+|\psi(1)|^2=1
\end{equation}
obeys the estimate

\begin{equation}\label{4:solns}
\| \psi \|_L \ge C_E L^{\gamma}
\end{equation}
for $L>0$ sufficiently large and some $E$-dependent constant $C_E$. Then $\Lambda$ is absolutely continuous with respect to $\mu_{\log b}$ for $b = 2 \gamma$.
\end{theorem}

\noindent{\it Remarks.} 1. The norm $\| \cdot \|_L$ in \eqref{4:solns} is defined as in \cite{dkl,jl1}, that is,
$$
\| \psi \|_L = \left( \sum_{n=0}^{\lfloor L \rfloor} \big| \psi(n) \big|^2 \; + \;(L-\lfloor L \rfloor) \big| \psi (\lfloor L \rfloor +1) \big|^2 \right)^{1/2}.
$$
2. This theorem and its proof have quantum dynamical consequences which will be discussed at the end of the paper.\\[2mm]
3. Our proof shows that one can draw even stronger continuity conclusions if the constant $C_E$ in \eqref{4:solns} can be chosen uniformly in the energy.\\[2mm]
4. We do not need that $V$ is bounded. Exponential upper bounds on solutions of \eqref{eve} are sufficient. This of course holds for bounded $V$ and since we have an application of Theorem~\ref{crit} to Sturmian potentials (which are bounded) in mind, we give the theorem in this simplified form.

\bigskip

The organization is as follows. We will prove Theorem~\ref{crit} in Section~2 and then apply it to Sturmian potentials in Section~3. Quantum dynamical applications are discussed in Section~4.

\section{Log-continuity of whole-line spectral measures}

The proof of Theorem~\ref{crit} follows a strategy similar to the one used in \cite{dkl}. Namely, we will first deduce $m$-function properties on a half-line from the assumptions on solutions to \eqref{eve}, uniformly in the boundary condition. In a second step we use the maximum modulus principle to show a similar property for the whole-line $m$-function which then implies the assertion of the theorem.

\begin{prop}\label{supest}
Fix $E\in\RR$. Suppose every solution of \eqref{eve} with \eqref{normal} obeys the estimate

\begin{equation}\label{upplow}
C_1 L^{\gamma} \leq \| \psi \|_L \leq C_2^L
\end{equation}
for constants $\gamma , C_1 , C_2$ and for $L>0$ sufficiently large. Then there exists $C_3$ such that for $\epsilon > 0$ small enough,

\begin{equation}\label{4:mEst}
\sup_\varphi \left| \frac{\sin(\varphi)+\cos(\varphi)m^+(E+i\epsilon)} {\cos(\varphi)-\sin(\varphi)m^+(E+i\epsilon)} \right| \leq C_3 \frac{\log b (\varepsilon)}{\epsilon},
\end{equation}
where $b=2\gamma$.
\end{prop}

\begin{proof}
Denote
$$
m_\varphi^+ (E + i \epsilon) = \frac{\sin(\varphi)+\cos(\varphi)m^+(E+i\epsilon)} {\cos(\varphi)-\sin(\varphi)m^+(E+i\epsilon)}
$$
and let $\psi_\varphi^{1/2}$ be the solutions to \eqref{eve} with
$$
\psi_\varphi^1 (0) = \sin(\varphi), \; \psi_\varphi^1 (1) = \cos(\varphi)
$$
and
$$
\psi_\varphi^2 (0) = -\cos(\varphi), \; \psi_\varphi^2 (1) = \sin(\varphi).
$$
Given $\epsilon > 0$, let $L_\varphi(\epsilon) > 0$ be defined by
$$
\| \psi_\varphi^1 \|_{L_\varphi(\epsilon)} \| \psi_\varphi^2 \|_{L_\varphi(\epsilon)} = \frac{1}{2 \epsilon}.
$$
Then the Jitomirskaya-Last inequality \cite{jl1} reads

\begin{equation}\label{jli}
\frac{5 - \sqrt{24}}{| m_\varphi^+ (E + i \epsilon) |} < \frac{\| \psi_\varphi^1 \|_{L_\varphi(\epsilon)}}{\| \psi_\varphi^2 \|_{L_\varphi(\epsilon)}} < \frac{5 + \sqrt{24}}{| m_\varphi^+ (E + i \epsilon) |}.
\end{equation}

From \eqref{upplow} and \eqref{jli} we get for $\epsilon > 0$ small enough,

\begin{eqnarray*}
\frac{\epsilon}{\log b (\epsilon)} \sup_\varphi | m_\varphi^+ (E + i \epsilon) |
& \le & \frac{\epsilon}{\log b (\epsilon)} \sup_\varphi (5 + \sqrt{24}) \frac{\| \psi_\varphi^2 \|_{L_\varphi(\epsilon)}}{\| \psi_\varphi^1 \|_{L_\varphi(\epsilon)}}\\
& = & \sup_\varphi (5 + \sqrt{24}) \frac{\| \psi_\varphi^2 \|_{L_\varphi(\epsilon)}}{\| \psi_\varphi^1 \|_{L_\varphi(\epsilon)}} \times \\
&& \times \frac{1}{2 \| \psi_\varphi^1 \|_{L_\varphi(\epsilon)} \| \psi_\varphi^2 \|_{L_\varphi(\epsilon)}}  \frac{1}{\log b (\frac{1}{2 \| \psi_\varphi^1 \|_{L_\varphi(\epsilon)} \| \psi_\varphi^2 \|_{L_\varphi(\epsilon)}})}\\
& = & \frac{(5 + \sqrt{24})}{2} \sup_\varphi \frac{\log (2 \| \psi_\varphi^1 \|_{L_\varphi(\epsilon)} \| \psi_\varphi^2 \|_{L_\varphi(\epsilon)})^b}{\| \psi_\varphi^1 \|_{L_\varphi(\epsilon)}^2}\\
& \le & \frac{(5 + \sqrt{24})}{2} \sup_\varphi \frac{\log (2 C_2^{2 L_\varphi (\epsilon)})^b}{C_1^2 L_\varphi (\epsilon)^{2 \gamma}}\\
& \le & C_3
\end{eqnarray*}
if $b = 2 \gamma$.
\end{proof}

\begin{prop}\label{piac}
Given a Borel set $\Sigma$, suppose that the estimate \eqref{upplow} holds for every $E \in \sigma(H)$ with $C_1,C_2$ independent of $E$. Then, given any function $m^-\!:\!\CC^+\to\CC^+$, and any $E \in \Sigma$,

\begin{equation}\label{m:n}
|m(E+i\epsilon)| = \left| \frac{m^+(E+i\epsilon)+m^-(E+i\epsilon)}{1-m^+(E+i\epsilon)m^-(E+i\epsilon)} \right| \leq C_3 \frac{\log b (\varepsilon)}{\epsilon}
\end{equation}
for all $\epsilon > 0$, where $b = 2 \gamma$. Consequently, $\Lambda(E)$ is uniformly $\log b$-Lipschitz continuous at all points $E \in \Sigma$. In particular, $\Lambda$ is absolutely continuous with respect to $\mu_{\log b}$ on $\Sigma$.

If \eqref{upplow} holds only with $E$-dependent constants $C_1, C_2$, but with a uniform $\gamma$, we can still deduce absolute continuity of $\Lambda$ with respect to $\mu_{\log b}$.
\end{prop}

\begin{proof}
Fix $E \in \Sigma$ and $\epsilon > 0$. Let $z=e^{2i\varphi}$ and $\mu=(m^+-i)/(m^++i)$. We may then rewrite \eqref{4:mEst} as
$$
\sup_{|z|=1} \left| \frac{1+\mu z}{1-\mu z} \right| \leq C_3 \frac{\log b (\varepsilon)}{\epsilon}.
$$
By $\Im(m^+) > 0$ we have $| \mu |<1$ and so $(1+\mu z)/(1-\mu z)$ defines an analytic function on $\{z : |z|\leq 1\}$. The point $z=(i-m^-)/(i+m^-)$ lies inside the unit disk since $\Im(m^-) > 0$. We have
$$
m = i \cdot \frac{1 + \mu \left( \frac{i - m^-}{i + m^-} \right) }{1 - \mu \left( \frac{i - m^-}{i + m^-} \right) }
$$
as can be checked by direct calculation. The estimate \eqref{m:n} thus follows from the maximum modulus principle. This estimate and the representation \eqref{4:mRep} provide
$$
\Lambda\big([E-\epsilon,E+\epsilon]\big) \leq 2 \epsilon \, \Im \big(m(E+i\epsilon)\big)\leq 2C_3\,\log b (\varepsilon) \quad \text{for all $E\in\Sigma$, $\epsilon>0$,}
$$
from which $\Lambda(E)$ is uniformly $\log b$-Lipschitz continuous
on $\Sigma$.

If we permit $C_1,C_2$ to depend on $E$, the only consequence is
that now $C_3$ depends on $E$ and so $\Lambda$ need not be
\textit{uniformly} Lipschitz continuous. However, absolute
continuity is still guaranteed.
\end{proof}

\begin{proof}[\textit{Proof of Theorem~\ref{crit}.}]
The assertion follows from Propositions~\ref{supest} and \ref{piac}.
\end{proof}

\section{Application to Sturmian Potentials}

In this section we discuss the case where $V$ is given by
\begin{equation}\label{sturmpot}
V(n) = \lambda \chi_{[1-\theta , 1)} (n \theta + \beta \mod 1).
\end{equation}
The non-trivial situation (i.e., non-periodic) is when we assume
that the coupling constant $\lambda$ is nonzero and the rotation
number $\theta \in (0,1)$ is irrational. Operators $H$ with such
potentials are standard models for one-dimensional quasicrystals;
see, for example, \cite{bist,d2}. It is quite easy to see that the
spectrum of $H$ does not depend on $\beta$ and can hence be
denoted by $\Sigma_{\lambda,\theta}$. It is known that for all
parameter choices, subject to the above conditions, the operator
$H$ has purely singular continuous spectrum \cite{bist,dkl}. More
detailed studies of the singular continuous spectral type can be
found in \cite{d1,dkl,jl2} where absolute continuity with respect
to $x^\alpha$ Hausdorff measures is established for certain
parameter values. Explicitly, it is known that for every
$\lambda$ and every bounded density number $\theta$, there is
$\alpha > 0$ such that for every $\theta$, the spectral measures
of $H$ are absolutely continuous with respect to $\mu_{x^\alpha}$.
Recall that $\theta$ is called a bounded density number if the
coefficients $a_n$ in the continued fraction expansion of
$\theta$,
$$
\theta = \cfrac{1}{a_1+ \cfrac{1}{a_2+ \cfrac{1}{a_3 + \cdots}}}
$$
satisfy $\limsup_{n \rightarrow \infty} \frac{1}{n} \sum_{i = 1}^n a_i < \infty$. The set of bounded density numbers is small in Lebesgue sense: it has measure zero. Let us define the associated rational approximants $\frac{p_n}{q_n}$ of $\theta$ by

\begin{alignat*}{3}
p_0 &= 0, &\quad    p_1 &= 1,   &\quad  p_n &= a_n p_{n-1} + p_{n-2},\\
q_0 &= 1, &     q_1 &= a_1, &       q_n &= a_n q_{n-1} + q_{n-2}.
\end{alignat*}
Our goal here is to establish absolute continuity of spectral measures with respect to logarithmic Hausdorff measures for \textit{almost every} $\theta$.

\begin{theorem}\label{sturmappl}
For every $\lambda$ and almost every $\theta$, there is $b > 0$ such that for every $\beta$, $\Lambda$ is absolutely continuous with respect to $\mu_{\log b}$.
\end{theorem}

This theorem follows from Theorem~\ref{crit} and the following proposition from \cite{dkl}:

\begin{prop}\label{lowerpower}
Let $\theta$ be such that for some $B < \infty$, $q_n \le B^n$ for every $n \in \NN$. Then for every $\lambda$, there exist $0 < \gamma, C < \infty$ such that for every $E \in \Sigma_{\lambda,\theta}$ and every $\beta$, every normalized solution $u$ of \eqref{eve} obeys

\begin{equation}\label{lpb}
\|u\|_L \ge C L^{\gamma}
\end{equation}
for $L$ sufficiently large.
\end{prop}

Note that the assumption of this proposition is obeyed by almost every $\theta$ \cite{khin}.

\section{Dynamical Implications}

Continuity properties of spectral measures imply quantum dynamical
bounds, as demonstrated by works of Guarneri \cite{g3}, Combes
\cite{c}, and Last \cite{l}; among others. In particular, Last
derives dynamical bounds from the non-singularity of spectral
measures with respect to $x^\alpha$--Hausdorff measures. More
recently, Landrigan \cite{l2} has observed that these results hold
for general Hausdorff measures, including the logarithmic
Hausdorff measures which are of primary interest in the present
article. Let us briefly recall the results of \cite{l2} and
discuss their consequences for Sturmian models.

Let $H$ be as in \eqref{oper} and let $\phi \in \ell^2(\ZZ)$. The Schr\"odinger time evolution is given by $\phi (t) = e^{-itH} \phi$ and the ``spreading'' of $\phi (t)$ is usually measured by considering two quantities, the survival probability
$$
| \langle \phi , \phi (t) \rangle |^2 = \left| \int e^{-itx} d \mu_\phi (x) \right|^2 = | \hat{\mu}_\phi (t) |^2,
$$
where $\mu_\phi$ is the spectral measure corresponding to the pair $(H, \phi)$, and expectation values
$$
\langle |X|^m \rangle (t) = \langle e^{-itH} \phi, |X|^m e^{-itH} \phi \rangle
$$
of moments of the position operator
$$
|X|^m = \sum_{n \in \ZZ} |n|^m \langle \delta_n , \cdot \rangle \delta_n.
$$
In the case of singular continuous spectral measures it is natural to consider time averaged quantities. Then, intuitively, the faster the spreading, the faster the decay of $\langle | \hat{\mu}_\phi |^2 \rangle_T$ and the faster the increase of $\langle \langle |X|^m \rangle \rangle_T$, where the time average $\langle \cdot \rangle_T$ is defined by $\langle f \rangle_T = \frac{1}{T} \int_0^T f(t) dt$. These relations are made explicit by the following pair of propositions which are simplified versions of Lemma~12 and Theorem~6 of \cite{l2}, respectively. In particular, they show that the dynamical bound one can prove is naturally related to the maximal dimension function one can pick to get a desired continuity property.

\begin{prop}\label{bound1}
If $\mu_\phi$ is uniformly $h$--Lipschitz continuous, then there is $C_\phi > 0$ such that
$$
\langle | \hat{\mu}_\phi |^2 \rangle_T < C_\phi \cdot h \left( \frac{1}{T} \right).
$$
\end{prop}

\begin{prop}\label{bound2}
If $\mu_\phi$ is absolutely continuous with respect to the Hausdorff measure $\mu_h$, then for each $m > 0$, there is $D_{\phi,m} > 0$ such that
$$
\langle \langle |X|^m \rangle \rangle_T > D_{\phi,m} \cdot h \left( \frac{1}{T} \right)^{-m}.
$$
\end{prop}

\noindent\textit{Remark.} The assumption can be relaxed. It suffices that $\mu_\phi$ is not singular with respect to $\mu_h$.

\bigskip

Let us now state the dynamical bounds we obtain for Sturmian potentials. Note that since $\{\delta_0,\delta_1\}$ is cyclic for $H$, absolute continuity of $\Lambda$ with respect to $\mu_h$ is inherited by $\mu_\phi$ for \textit{all} $\phi \in \ell^2(\ZZ)$ and uniform $h$-Lipschitz continuity is inherited by $\mu_\phi$ for all $\phi \in \ell^2(\ZZ)$ of \textit{compact support}.

\begin{coro}
For every $\lambda$ and almost every $\theta$, there is $b > 0$ such that for every $\beta$, the Sturmian operator corresponding to the triple $(\lambda, \theta, \beta)$ satisfies the following:\\[2mm]
{\rm (a)} For every $\phi \in \ell^2(\ZZ)$, there is $C_\phi > 0$ such that
$$
\langle | \hat{\mu}_\phi |^2 \rangle_T < C_\phi \cdot (\log T)^{-b}.
$$
{\rm (b)} For every $\phi \in \ell^2(\ZZ)$ of compact support and every $m > 0$, there is $D_{\phi,m}$ such that
$$
\langle \langle |X|^m \rangle \rangle_T > D_{\phi,m} \cdot (\log T)^{bm}.
$$
\end{coro}

\begin{proof}
This follows from Theorem~\ref{sturmappl} along with Propositions~\ref{bound1} and \ref{bound2}.
\end{proof}

\end{document}